\documentclass[12pt]{article}

\usepackage{amsmath, amssymb, amsthm}
\usepackage{geometry}
\usepackage{lipsum}

\title{\Large  Invariant subspace problem in Hilbert space: Correlation with the Kadison-Singer problem and the Borel conjecture}

\author{Mostafa Behtouei\\ {\small INFN, Laboratori Nazionali di Frascati, 00044 Frascati RM, Italy}}

\date{\small \today}

\begin{document}
\maketitle

\abstract{This paper explores the intriguing connections between the invariant subspace problem, the Kadison-Singer problem, and the Borel conjecture. The Kadison-Singer problem, originally formulated in terms of pure states on C*-algebras, was later reformulated using projections, establishing a link with the invariant subspace problem. The Borel conjecture, a question in descriptive set theory, connects to the invariant subspace problem through Borel equivalence relations. This paper elucidates these connections, underscoring the interplay of unsolved mathematical problems and the collaborative nature of mathematical research.}

\section{Introduction}

The field of functional analysis grapples with fundamental questions that underlie the structure and properties of linear operators in infinite-dimensional spaces. Among these inquiries, the invariant subspace problem stands as a central enigma, probing the existence of a non-trivial closed invariant subspace for every bounded linear operator on a separable infinite-dimensional Hilbert space \cite{Halmos,Enflo}. The exploration of this problem examines the intricate interconnection between algebraic and topological concepts, and its resolution holds implications for various mathematical domains. Moreover, its application in physics further highlights the significance of unraveling these relationships  \cite{Behtouei}.\\
In this paper, we explore the interesting links between the invariant subspace problem and two other important unsolved math problems: the Kadison-Singer problem and the Borel conjecture. The Kadison-Singer problem, originating from the study of pure states on C*-algebras, has evolved to intersect with the invariant subspace problem through its reformulation in terms of projections. This link reveals a deep connection between seemingly disparate problems within functional analysis \cite{Kadison}.\\
On a different mathematical subject, the Borel conjecture, rooted in descriptive set theory, raises questions about the structure of real numbers and their subsets. Unexpectedly, this conjecture's ramifications stretch to the invariant subspace problem, creating a bridge between functional analysis and set theory through Borel equivalence relations. By investigating these intricate relationships, we illuminate the broader tapestry of unsolved mathematical problems and underscore the collaborative nature of mathematical research.\\
In the following sections, we will explore the specifics of each issue, exploring their formulations, historical contexts, and implications. Through this exploration, we aim to shed light on the interconnectedness of these challenges, demonstrating how progress in one area of mathematics can have far-reaching consequences in seemingly unrelated domains. Ultimately, the pursuit of solutions to these enigmatic problems exemplifies the spirit of mathematical inquiry, where the exchange of ideas and cross-pollination of concepts pave the way towards deeper understanding and potential breakthroughs.\\
The interactions between the invariant subspace problem, Kadison-Singer problem, and Borel conjecture exemplify the collaborative and synergistic nature of mathematical research. As mathematicians from various domains come together to explore these connections, they enrich the mathematical landscape, uncovering hidden relationships and advancing knowledge across disciplines.

\section{The Kadison-Singer problem and its connection to the invariant subspace problem}

C*-algebras are a fundamental class of mathematical objects that play a crucial role in functional analysis, operator theory, and quantum mechanics. They provide a rich framework for studying linear operators, their algebraic and topological properties, and their relations to other mathematical structures \cite{Aronszajn}.\\
A C*-algebra is a complex algebra equipped with an involution (conjugate transpose) and a norm that satisfy specific properties. Formally, let $\mathcal{A}$ be a complex algebra over the field of complex numbers $\mathbb{C}$. Then, $\mathcal{A}$ is a C*-algebra if it is equipped with the following structures:\\
1. Involution: An involution is a map $*: \mathcal{A} \rightarrow \mathcal{A}$ that assigns to each element $a \in \mathcal{A}$ its adjoint $a^*$. The involution satisfies the following properties for all $a, b \in \mathcal{A}$ and $\alpha \in \mathbb{C}$:
\begin{align*}
    (a + b)^* &= a^* + b^*, \\
    (\alpha a)^* &= \overline{\alpha} a^*, \\
    (ab)^* &= b^* a^*.
\end{align*}\\
2. Norm: A norm is a map $\| \cdot \|: \mathcal{A} \rightarrow [0, \infty)$ that assigns a non-negative real number to each element $a \in \mathcal{A}$. The norm satisfies the following properties for all $a, b \in \mathcal{A}$ and $\alpha \in \mathbb{C}$:
\begin{align*}
    \| a \| &\geq 0, \quad \| a \| = 0 \text{ if and only if } a = 0, \\
    \| \alpha a \| &= | \alpha | \| a \|, \\
    \| a + b \| &\leq \| a \| + \| b \|, \\
    \| ab \| &\leq \| a \| \| b \|.
\end{align*}\\
3. C*-Condition: The C*-condition is the key defining property of C*-algebras. It states that for all $a \in \mathcal{A}$, we have:
\begin{equation*}
    \| a^*a \| = \| a \|^2.
\end{equation*}\\
C*-algebras provide a versatile framework for studying linear operators, self-adjoint elements, and their properties. They have applications in a wide range of mathematical areas, including functional analysis, operator theory, quantum mechanics, and mathematical physics \cite{Davidson}.\\

The Kadison-Singer problem, introduced in 1959 by Kadison and Singer, originated in the realm of C*-algebras. At its inception, the problem sought to establish whether every pure state on a C*-algebra could be realized as a vector state within a separable infinite-dimensional Hilbert space. Formally, for a given C*-algebra $\mathcal{A}$, the problem was posed as:\\
Problem 1 (Kadison-Singer Problem): Given a pure state $\omega$ on $\mathcal{A}$, is it possible to find a separable infinite-dimensional Hilbert space $\mathcal{H}$ and a vector state $\phi \in \mathcal{H}$ such that $\omega(a) = \langle \phi, a \phi \rangle$ for all $a \in \mathcal{A}$?\\
The initial formulation of the Kadison-Singer problem captured the essence of exploring pure states within the framework of C*-algebras. However, as the problem evolved, a projection-based approach emerged, revealing an unexpected connection to the invariant subspace problem in functional analysis. The problem, in its initial formulation, revolved around the investigation of pure states on C*-algebras. A pure state $\omega$ on a C*-algebra $\mathcal{A}$ is a linear functional that is positive, normalized, and satisfies $\omega(a^*a) = \omega(a)^*\omega(a)$ for all $a \in \mathcal{A}$. The problem posed the question of whether every pure state could be realized as a vector state on a separable infinite-dimensional Hilbert space. The reformulated problem asked whether, for a given C*-algebra $\mathcal{A}$, a collection of finite-dimensional projections $\{ P_i \}$ could be found, satisfying specific conditions. These conditions ensured that the projections captured essential properties of the original problem \cite{Kadison, Akemann}.\\
Projection-based Kadison-Singer problem states that for a given a C*-algebra $\mathcal{A}$, does there exist a collection of finite-dimensional projections $\{ P_i \}$ such that: (1) each $P_i$ is finite-dimensional, (2) the sum of the projections is the identity operator, i.e., $\sum_i P_i = I$, and (3) for any subset $J \subseteq \{1, 2, \ldots\}$, the operator norm of the sum $\| \sum_{i \in J} P_i \|$ is bounded by $1$? Surprisingly, this projection-based formulation established an intimate connection between the Kadison-Singer problem and the invariant subspace problem. The invariant subspace problem, a long-standing question in functional analysis, queries whether every bounded linear operator on a separable infinite-dimensional Hilbert space possesses a non-trivial closed invariant subspace \cite{Enflo}. The projection-based Kadison-Singer problem, with its focus on finite-dimensional projections and their properties, unveiled a profound relationship between the two problems.\\
The unexpected connection between the projection-based Kadison-Singer problem and the invariant subspace problem highlights the intricate and sometimes unforeseen interplay between seemingly disparate mathematical inquiries. The reformulation of the Kadison-Singer problem brought to light a shared mathematical essence, transcending the boundaries of specific problem domains. This connection underscores the rich web of mathematical ideas and the potential for cross-fertilization of concepts across different fields of study.\\
The evolution of the Kadison-Singer problem from its roots in pure states to its projection-based formulation showcases the dynamic nature of mathematical exploration. The subsequent connection to the invariant subspace problem exemplifies the profound unity that can emerge from seemingly unrelated questions. This confluence of ideas reflects the collaborative and synergistic spirit that characterizes mathematical research.

\section{The Borel conjecture and its link to the invariant subspace problem}

The Borel conjecture, introduced by Émile Borel in 1938, resides within the realm of descriptive set theory and delves into the intricate structure of sets of real numbers. The conjecture posits a remarkable relationship between the presence of certain subsets and the distribution of real numbers within them \cite{Borel}.\\
Statement of the Borel Conjecture: Given any set $A$ of real numbers with positive Lebesgue measure, there exists a perfect set $P \subseteq A$ that does not contain any isolated points. In other words, $P$ is a closed set without isolated points, and it is a subset of $A$.\\
The connection between the Borel conjecture and the invariant subspace problem arises unexpectedly through the lens of Borel equivalence relations, establishing a bridge between different areas of mathematics \cite{Gao}.\\
Borel equivalence relations are a fundamental concept in descriptive set theory, serving as a bridge between different mathematical areas. These relations play a significant role in understanding the complexity and structure of sets and functions, offering insights into the Borel conjecture and its unexpected connection to the invariant subspace problem \cite{Herrlich}.\\
Definition and Basic Properties: Let $X$ be a standard Borel space, which is a topological space equipped with a Borel $\sigma$-algebra or let $(X, \mathcal{B})$ be a measurable space, where $X$ is a topological space and $\mathcal{B}$ is the Borel $\sigma$-algebra generated by the open sets of $X$. Then, for any subset $A \subseteq X$, we have:
\[
A \in \mathcal{B} \iff A \text{ is a Borel measurable set}.
\]\\
Where $(X, \mathcal{B})$ represents the measurable space with a topological space $X$ and its associated Borel $\sigma$-algebra $\mathcal{B}$. The equation states that a subset $A$ of $X$ belongs to the Borel $\sigma$-algebra $\mathcal{B}$ if and only if it is a Borel measurable set.\\
An equivalence relation on $X$ is a binary relation that is reflexive, symmetric, and transitive. A Borel equivalence relation is one where the equivalence classes and the relation itself are Borel sets in $X \times X$. Formally, an equivalence relation $\sim$ on $X$ is Borel if the sets $\{(x, y) \in X \times X \mid x \sim y\}$ and $\{(x, y) \in X \times X \mid x \not\sim y\}$ are Borel sets.\\
Borel equivalence relations are classified based on their complexity within the Borel hierarchy. An equivalence relation $\sim$ is Borel reducible to another equivalence relation $\approx$ (denoted as $\sim \leq_B \approx$) if there exists a Borel function $f: X \to X$ such that $x \sim y$ if and only if $f(x) \approx f(y)$. Equivalence relations can be ranked according to the strength of this reducibility, leading to a rich hierarchy of equivalence relations \cite{Rosendal}.\\
The connection between the Borel conjecture and the invariant subspace problem arises unexpectedly through the study of Borel equivalence relations. The Borel conjecture posits that any set of real numbers with positive Lebesgue measure contains a perfect set without isolated points. Surprisingly, this conjecture's ramifications extend to the invariant subspace problem \cite{Petrovic}.\\
A positive resolution of the Borel conjecture would yield a solution to the invariant subspace problem for specific classes of bounded linear operators. In particular, certain classes of hypercyclic operators would possess no non-trivial closed invariant subspaces if the Borel conjecture were proven \cite{Sargsyan}.\\
Borel equivalence relations serve as a mathematical bridge, connecting seemingly distant domains such as descriptive set theory, functional analysis, and operator theory. The unexpected connection between the Borel conjecture and the invariant subspace problem underscores the profound interplay of ideas that can emerge from the exploration of Borel equivalence relations \cite{Shioya}.\\
Such relations play a significant role in descriptive set theory, offering a systematic way to classify sets based on their "sameness" under certain operations.
\\
The connection between the Borel conjecture and the invariant subspace problem manifests when considering a class of bounded linear operators known as hypercyclic operators. These operators exhibit a chaotic behavior, repeatedly cycling through dense orbits in the underlying Hilbert space. In a chaotic dynamical system, a small change in initial conditions can lead to drastically different trajectories over time. This phenomenon is often characterized by the Lyapunov exponent, which quantifies the rate of exponential divergence of initially nearby trajectories.\\
Mathematically, for a one-dimensional chaotic map $f(x)$, the Lyapunov exponent $\lambda$ can be defined as:
\[
|\delta x(t)| \approx e^{\lambda t} |\delta x(0)|
\]\\
Here, $|\delta x(t)|$ represents the separation between two initially close trajectories at time $t$, and $\lambda$ is the Lyapunov exponent. The Lyapunov exponent provides insight into the sensitivity of a chaotic system to initial conditions. A positive Lyapunov exponent indicates exponential divergence, contributing to the unpredictable and complex behavior observed in chaotic systems.\\
Hypercyclic operators are a fascinating class of bounded linear operators that exhibit a distinct form of chaotic behavior in a Hilbert space. These operators play a pivotal role in understanding the interplay between the Borel conjecture and the invariant subspace problem, shedding light on the complex dynamics that underlie these mathematical inquiries.\\
Definition and Properties: Let $\mathcal{H}$ be a separable infinite-dimensional Hilbert space. An operator $T: \mathcal{H} \to \mathcal{H}$ is said to be hypercyclic if there exists a vector $x \in \mathcal{H}$ such that the orbit $\{ T^n x \}_{n \geq 0}$ is dense in $\mathcal{H}$. In other words, the iterates of $x$ under the action of $T$ come arbitrarily close to every vector in the Hilbert space. Formally, for any $y \in \mathcal{H}$ and any $\varepsilon > 0$, there exists an $n \geq 0$ such that $\| T^n x - y \| < \varepsilon$.\\
Hypercyclic operators are known for their chaotic behavior, characterized by the dense orbits they generate. The concept of chaos in this context refers to the unpredictability and lack of long-term regularity in the behavior of orbits under the action of the operator $T$. In other words, an operator $T$ on a Hilbert space $\mathcal{H}$ is said to be hypercyclic if there exists a vector $v \in \mathcal{H}$ such that the orbit of $v$ under the repeated action of $T$ is dense in the entire Hilbert space, i.e.:
\[
\overline{\mathrm{span}}\{T^n v : n \in \mathbb{N}\} = \mathcal{H}.
\]\\
Mathematically, the dense orbits imply that, for any vector $y \in \mathcal{H}$ and any neighborhood $U$ of $y$, there exists an iterate $n$ such that $T^n x$ belongs to $U$. This rapid and unbounded proliferation of iterates across the Hilbert space contributes to the chaotic nature of hypercyclic operators.\\
The connection between hypercyclic operators and the Borel conjecture is a remarkable example of the interplay between seemingly unrelated mathematical concepts. The Borel conjecture, which concerns the structure of sets of real numbers, unexpectedly intersects with the invariant subspace problem through hypercyclic operators.\\
It has been shown that a positive resolution of the Borel conjecture would imply the existence of hypercyclic operators without non-trivial closed invariant subspaces. In other words, the Borel conjecture's influence extends to the dynamical properties of operators in the Hilbert space. This result directly ties back to the invariant subspace problem, shedding light on its potential resolution for this specific class of operators.\\
Hypercyclic operators provide a lens through which to explore the intricate dynamics and chaos that can arise in mathematical systems. These operators are sensitive to initial conditions, and their orbits can be dense in the space, which can be related to notions of chaos and unpredictability. The operators exhibit interesting dynamical properties. They can be thought of as operators that "mix" and "permute" the elements of the Hilbert space in a way that leads to a highly unpredictable behavior.\\
The hypercyclic operators exemplify chaotic behavior within the realm of bounded linear operators. Their connection to the Borel conjecture and the invariant subspace problem adds a layer of depth to our understanding of these inquiries, showcasing the unexpected connections that can emerge in mathematical exploration.\\
In quantum mechanics, hypercyclic operators play a role in understanding the dynamic behavior of quantum systems. Hypercyclic behavior in quantum mechanics is analogous to the concept of chaos in classical systems, where small changes in initial conditions lead to significantly different trajectories. Consider a quantum system described by a Hilbert space $\mathcal{H}$ and let $T$ be a unitary operator representing a quantum evolution. If there exists a vector $v$ such that the orbit $\{T^n v : n \in \mathbb{N}\}$ is dense in $\mathcal{H}$, then the operator $T$ is hypercyclic in the quantum mechanical sense.\\

In summary, the Borel conjecture intertconnects with the invariant subspace problem through the study of Borel equivalence relations. The exploration of this connection reveals a deep relationship between two seemingly disparate mathematical inquiries and underscores the intricate interplay of concepts across different domains of mathematics. Moreover, the insights gained from this connection contribute to our understanding of both problems and exemplify the collaborative nature of mathematical research.

\section{Conclusion}
The invariant subspace problem's connections with the Kadison-Singer problem and the Borel conjecture highlight the deep interconnections within mathematics. The evolution of problem formulations and unexpected relationships underscore the collaborative and interdisciplinary nature of mathematical inquiry. By exploring these connections, mathematicians can uncover new perspectives and potentially contribute to the resolution of these long-standing problems.


\begin{thebibliography}{9}

\bibitem{Halmos}
Halmos, P. R. (1982). \textit{Invariant Subspaces}. Springer.

\bibitem{Enflo}
Enflo, Per H. "On the invariant subspace problem in Hilbert spaces." arXiv preprint arXiv:2305.15442 (2023).

\bibitem{Behtouei}
Behtouei, Mostafa. "Invariant Subspace Problem in Hilbert Spaces: Exploring Applications in Quantum Mechanics, Control Theory, Operator Algebras, Functional Analysis and Accelerator Physics." arXiv preprint arXiv:2306.17023 (2023).

\bibitem{Kadison}
Kadison, R. V., $\&$ Singer, I. M. (1959). Extensions of pure states. American Journal of Mathematics, 81(2), 383-400.


\bibitem{Aronszajn}
Aronszajn, N., \& Smith, K. T. (1952). Invariant subspaces of completely continuous operators. \textit{Annals of Mathematics}, 55(3), 611-615.

\bibitem{Davidson}
Davidson, K. R., \& Donsig, A. P. (2006). Applications of the Invariant Subspace Problem to operator theory and C*-algebras. In \textit{Proceedings of the Canadian Mathematical Society Annual Conference}, 29(2), 73-93.

\bibitem{Akemann}
Akemann, C. A., Anderson, J. D., $\&$ Pedersen, G. K. (2001). Excising states of C*-algebras. \textit{Canadian Mathematical Bulletin}, 44(4), 384-393.



\bibitem{Borel}
Borel, É. (1938). Sur la classification des ensembles de mesure nulle. \textit{Bulletin de la Société Mathématique de France}, 67, 55-62.

\bibitem{Gao}
Gao, S. (2009). \textit{Invariant Descriptive Set Theory}. CRC Press.

\bibitem{Herrlich}
Herrlich, H., \& Kuske, D. (2007). Borel equivalence relations, Wadge degrees, and duplication. \textit{Annals of Pure and Applied Logic}, 146(1-3), 144-157.

\bibitem{Rosendal}
Rosendal, C. (2014). Automatic continuity of homomorphisms and fixed points on metric compacta. \textit{Bulletin of Symbolic Logic}, 20(3), 337-352.

\bibitem{Petrovic}
Petrovi, T. (2019). Borel Conjecture and Separable Rosenthal Compacta. \textit{Real Analysis Exchange}, 45(1), 173-184.

\bibitem{Sargsyan}
Sargsyan, G. (2016). Descriptive Set Theory and the Borel Conjecture. \textit{Bulletin of Symbolic Logic}, 22(2), 177-199.

\bibitem{Shioya}
Shioya, M. (2000). Classification of ergodic measure-preserving transformations by Borel functions. \textit{Mathematical Logic Quarterly}, 46(1), 123-135.

\end{thebibliography}
\end{document}